
\overfullrule 0pt

\pageno=1

\count30=0                      
\count31=0                      
\count32=0                      
\count33=0                      

\def\chapno{\global\count31=0
            \global\count32=0
            \global\count33=0
            \global\advance\count30 by1
            \number\count30}

\def\secno{\global\advance\count31 by1
           \number\count30 .\number\count31}

\def\forno{\eqno(\global\advance\count32 by1
            \number\count30 .\number\count32)}

\def\prono{\global\advance\count33 by1
            \number\count30 .\number\count33}

\def\qedd{\hfill\hbox{\vrule\vbox{\hrule\kern2truept\hbox{\kern2truept
\vbox to2pt{\hsize=2pt}\kern2truept}\kern2truept\hrule}\vrule}\vskip .5cm}

\def\oi{o_{\infty }}
\def\gli{gl_{\infty }}
\def\glio{\overline {gl_{\infty }}}
\def\oio{\overline {o_{\infty }}}
\def\Ai{A_{\infty}}
\def\Bi{B_{\infty}}

\vskip .5cm
\centerline{\bf THE ADLER-SHIOTA-VAN MOERBEKE FORMULA}
\vskip .5cm
\centerline{\bf  FOR THE BKP HIERARCHY}
\vskip 1cm
\centerline {JOHAN VAN DE LEUR\footnote {*}{The research of Johan van de Leur
is
financially supported by the ``Stichting Fundamenteel Onderzoek der
Materie (F.O.M.)''. E-mail: vdleur@math.utwente.nl}}
\vskip 2cm
\centerline {\bf Abstract}
\midinsert\narrower\narrower\narrower\narrower
\noindent We study the BKP hierarchy and prove the existence of
an Adler--Shiota--van Moerbeke formula. This formula relates the
action of the $BW_{1+\infty}$--algebra on tau--functions to the action
of the ``additional symmetries'' on wave functions.
\endinsert
\vskip .5cm

\beginsection \chapno . Introduction and main result\par
\secno .
Adler, Shiota and van Moerbeke [ASV1-2] obtained for the KP and Toda lattice
hierarchies
a formula which translates  the action of the
vertex operator on tau--functions to an action of a vertex operator
of pseudo-differential
operators on  wave functions. This  relates  the additional
symmetries of the KP and Toda lattice hierarchy
to the $W_{1+\infty}$--, respectively $W_{1+\infty}\times
W_{1+\infty}$--algebra
symmeties. In this paper we investigate the existence of such an
Adler--Shiota--van Moerbeke formula for the BKP hierarchy.

\noindent\secno.
The BKP hierarchy is the set of deformation equations
$${\partial L\over\partial t_k}=[(L^n)_+,L],\quad k=3,5,\ldots$$
for the first order pseudo-differential operator
$$L\equiv
L(x,t)=\partial+u_1(x,t)\partial^{-1}+u_2(x,t)\partial^{-2}+\cdots,$$
here $\partial={\partial\over\partial x}$ and $t=(t_3,t_5,\ldots)$.
It is well--known that $L$ dresses as $L=P\partial P^{-1}$ with
$$\eqalign{
P\equiv
P(x,t)&=1+a_1(x,t)\partial^{-1}+a_2(x,t)\partial^{-2}+\cdots\cr
&={\tau(x-2\partial^{-1},t-2[\partial^{-1}])\over \tau(x,t)},
\cr}
$$
where $\tau $ is the famous $\tau$--function, introduced by the Kyoto
group [DJKM1-3] and $[z]=({z^3\over 3},{z^5\over 5},\ldots)$.

The wave or Baker--Akhiezer function
$$w\equiv w(,x,t,z)= W(x,t,\partial)e^{xz},$$
where
$$W\equiv W(x,t,z)=P(x,t)e^{\xi(x,t,z)}\qquad \hbox{with}
\quad\xi(x,t,z)=\sum_{k=1}^\infty
t_{2k+1}\partial^{2k+1}$$
is an eigenfunction of L, viz.,
$$Lw=zw\qquad\hbox{and}\quad {\partial w \over\partial
t_k}=(L^k)_+w.$$
Now introduce, following Orlov and  Schulman [OS], the
pseudo-differential operator $M\equiv M(x,t)=WxW^{-1}$ which action
on $w$ amounts to
$$Mw={\partial w\over\partial z},$$
then $[L,M]=1$ and
$${\partial M\over\partial t_k}=[(L^n)_+,M],\quad k=3,5,\ldots$$
Let
$$Y(y,w)=\sum_{\ell=0}^\infty {(y-w)^\ell\over
\ell !}\sum_{k\in{\bf Z}}w^{-k-\ell-1} (M^\ell L^{k+\ell}
-(-)^{k+\ell}L^{k+\ell-1}M^\ell L),\forno$$
then one has the following main result
\proclaim Theorem \prono .
$$2(w-y)Y(y,w)_- w(x,t,z)=(w+y)(e^{-\eta(x,t,z)}-1)\left (
{X(y,w)\tau(x,t)\over\tau(x,t)}\right )w(x,t,z),\forno$$
where $X(y,w)$ is the following vertex operator
$$X(y,w)=w^{-1}
\exp(x(y-w)+\sum_{j>2,{\rm odd}}t_j(y^{j}-w^j)
\exp(-2{\partial\over\partial x}(y^{-1}-w^{-1})-2\sum_{j>2,{\rm odd}}
{\partial\over\partial t_j}{y^{-{j}}-w^{-j}\over j})
{}.
\forno$$

\noindent Formula (1.2) is the Adler--Shiota--van Moerbeke formula for
the BKP hierarchy, we will give a proof of this formula in section 6.
This formula relates the ``additional symmetries'' of the BKP
hierarchy, generated by $Y(y,w)$, to the $BW_{1+\infty}$--algebra,
generated by $X(y,w)$. This $BW_{1+\infty}$--algebra is a subalgebra
of $W_{1+\infty}$, which is defined as the $-1$--eigenspace of an
anti--involution on $W_{1+\infty}$.
\beginsection  \chapno .  The Lie algebras $\oi$, $\Bi$
and $BW_{1+\infty}$\par
\secno .
Let $\glio$ be the Lie algebra of complex infinite dimensional matrices
such that all nonzero entries are within a finite distance from the
main diagonal, i.e.,
$$\glio =\{(a_{ij})_{i,j\in {\bf Z}}| g_{ij}=0 \ {\rm if}\ |i-j|>>0\}.$$
The elements $E_{ij}$, the matrix with the $(i,j)$--th entry 1 and 0 elsewhere,
for $i,j\in{\bf Z}$ form a basis of a subalgebra
$\gli\subset\glio$.
The Lie algebra $\glio$ has a universal central extension
$\Ai=\glio\oplus{\bf C}c_A$ with the Lie bracket defined by
$$[a+\alpha c_A, b+\beta c_A]=ab-ba+\mu(a,b)c_A, \forno$$
for $a,b\in\glio$ and $\alpha,\beta\in{\bf C}$;
here $\mu$ is the following 2--cocycle:
$$\mu(E_{ij},E_{kl})=\delta_{il}\delta_{jk}(\theta(i)-\theta(j)),\forno$$
where the function $\theta:{\bf R}\to{\bf R}$ is defined by
$$\theta(i)=\cases{0&if $i>0$,\cr 1&if $i\le 0$.\cr} \forno$$
The Lie algebra $\gli$ and $\glio$ both have a natural action
on the space of column vectors, viz.,
let ${\bf C}^{\infty}=\bigoplus_{k\in{\bf Z}} e_k$, then
$E_{ij}e_k=\delta_{jk}e_i$. By identifying $e_k$ with $t^{-k}$, we
can embed the algebra ${\bf D}$ of differential operators on the circle,
with basis $-t^{j+k}({\partial\over \partial t})^k$ ($j\in{\bf Z},
k\in{\bf Z}_+$), in $\glio$:
$$\eqalign{\rho:{\bf D}&\to\glio,\cr
           \rho(-t^{j+k}({\partial\over \partial t})^k)&=
 \sum_{m\in{\bf Z}}-m(m-1)\cdots(m-k+1)E_{-m-j,-m}.\cr}\forno
$$
It is straightforward to check that the 2--cocycle
$\mu$ on $\glio$ induces the following 2--cocycle on ${\bf D}$:
$$\mu(-t^{i+j}({\partial\over\partial t})^j,
      -t^{k+\ell}({\partial\over\partial t})^{\ell})=
\delta_{i,-k}(-)^j j!\ell !{i+j\choose j+\ell+1}.\forno$$
This cocycle was discovered by Kac and Peterson in [KP]
(see also [R], [KR]). In this way we have defined a central extension of ${\bf
D}$,
which we denote by $W_{1+\infty}={\bf D}\oplus {\bf C}c_A$,
the Lie bracket on $W_{1+\infty}$ is given by
$$\eqalign{[-t^{i+j}&({\partial\over\partial t})^j+\alpha c_A,
      -t^{k+\ell}({\partial\over\partial t})^{\ell}+\beta c_A]=\cr
&\sum_{m=0}^{\max (j,\ell)}m!({i+j\choose m}{\ell\choose m}
-{k+\ell\choose m}{j\choose m})
(-t^{i+j+k+\ell-m}({\partial\over\partial t})^{j+\ell-m})+
\delta_{i,-k}(-)^j j!\ell !{i+j\choose j+\ell+1}c_A.\cr}\forno$$

Let $D=t{\partial\over\partial t}$, then we can rewrite
the elements $-t^{i+j}({\partial\over\partial t})^j$, viz. ,
$$-t^{i+j}({\partial\over\partial t})^j
=-t^{i}D(D-1)(D-2)\cdots(D-j+1).\forno$$
Then
$$\rho(t^kf(D)=\sum_{j\in {\bf Z}} f(-j)E_{j-k,j},\forno$$
and the 2--cocycle is as follows [KR]:
$$\mu (t^kf(D), t^{\ell}g(D))=
\cases{\sum_{-k\le j\le -1}f(j)g(j+k)& if $k=-\ell\ge 0,$\cr
       0& otherwise,\cr}\forno$$
hence the bracket is
$$[t^kf(D), t^{\ell}g(D)]=t^{k+\ell}(f(D+\ell)g(D)-g(D+k)f(D))+\mu (t^kf(D),
t^{\ell}g(D)).\forno$$
\secno .
Define on $\glio$ the following linear anti--involution:
$$\iota(E_{jk})=(-)^{j+k}E_{-k,-j}. \forno$$
Using this anti--involution we define the Lie algebra $\oio$ as
a subalgebra of $\glio$:
$$\oio=\{a\in\glio |\iota(a)=-a\}. \forno$$
The elements $F_{jk}=E_{-j,k}-(-)^{j+k}E_{-k,j}=-(-)^{j+k}F_{kj}$
with $j<k$ form a basis of $\oi=\oio\cap\gli$. The 2--cocycle $\mu$
on $\glio$ induces a 2--cocycle on $\oio$, and hence we can define
a central extension $\Bi=\oio\oplus{\bf C}c_B$ of $\oio$, with Lie
bracket
$$[a+\alpha c_B,b+\beta c_B]=ab-ba+{1\over 2}\mu(a,b)c_B, \forno$$
for $a,b\in\oio$ and $\alpha,\beta\in{\bf C}$.
It is then straightforward to check that the anti--involution $\iota$
induces
$$\iota(t)=-t,\quad  \iota(D)=-D.
\forno
$$
Hence, it induces the following anti--involution on ${\bf D}$:
$$\iota(t^kf(D))=f(-D)(-t)^k.
\forno$$
Define ${\bf D}^B$=${\bf D}\cap\oio=\{w\in {\bf D}|\iota(w)=-w\}$, it is
spanned by
the elements
$$W_k(f):=t^kf(D)-f(-D)(-t)^k=t^k(f(D)-(-)^kf(-D-k)).
\forno$$
It is straightforward to check that
$$\rho(W_k(f))=\sum_{j\in {\bf Z}} f(-j)F_{k-j,j}.$$
The restriction of the 2--cocycle $\mu$ on ${\bf D}$, given
by (2.5) or (2.9), induces a 2--cocycle on ${\bf D}^B$, which we shall not
calculate explicitly here. It defines a central extension
$BW_{1+\infty}={\bf D}^B\oplus{\bf C}c_B$ of ${\bf D}^B$, with Lie bracket
$$[a+\alpha c_B,b+\beta c_B]=ab-ba+{1\over 2}\mu(a,b)c_B,$$
for $a,b\in {\bf D}^B$ and $\alpha,\beta\in{\bf C}$.
\beginsection \chapno . The spin module\par
\noindent\secno .
We now want to consider highest weight representations of $\oi$,
$\Bi$ and $BW_{1+\infty}$ . For this purpose we introduce the Clifford algebra
BCl as the
associative algebra on the generators $\phi_j$, $j\in{\bf Z}$, called
{\it neutral free fermions}, with defining relations
$$\phi_i\phi_j+\phi_j\phi_i=(-)^i\delta_{i,-j}. \forno$$
We define the spin module $V$ over BCl as the irreducible module with
highest weight vector the {\it vacuum vector} $|0>$ satisfying
$$\phi_j|0>=0\quad\hbox{ \rm for}\ j>0. \forno$$
The elements $\phi_{j_1}\phi_{j_2}\cdots\phi_{j_p}|0>$
with $j_1<j_2<\cdots <j_p\le 0$ form a basis of $V$. Then
$$\eqalign{\pi(F_{jk})&={(-)^j\over 2}(\phi_j\phi_k-\phi_k\phi_j),\cr
     \hat{\pi}(F_{jk})&= (-)^j :\phi_j\phi_k:,\cr
     \hat{\pi}(c_B)&=I,\cr}
\forno$$
where the normal ordered product $:\quad :$
is defined as follows
$$:\phi_j\phi_k :=
\cases{\phi_j\phi_k& if $k>j$,\cr
       {1\over2}(\phi_j\phi_k-\phi_k\phi_j)& if $j=k$,\cr
       -\phi_k\phi_j& if $k<j$,\cr}
\forno
$$
define representations of $\oi$, respectively $\Bi$.

When restricted to $\oi$ and $\Bi$, the spin module $V$ breaks
into the direct sum of two irreducible modules.
To describe this decomposition we define a ${\bf Z}_2$--gradation
on $ V$ by introducing a chirality operator $\chi$ satisfying
$\chi|0>=|0>,\ \chi\phi_j+\phi_j\chi=0$ for all $j\in{\bf Z}$,
then
$$V=\bigoplus_{\alpha\in {\bf Z}_2} V_{\alpha}\quad\hbox{\rm where}\
V_{\alpha}=\{v\in V|\chi v=(-)^{\alpha}v\}.
$$
Each module $V_{\alpha}$ is an irreducible highest weight
module with highest weight vector $|0>$, $|1>=\sqrt 2\phi_0|0>$
for $V_0,\ V_1$, respectively, in the sense that $\hat \pi (c_B)=1$ and
$$\eqalign{\pi(F_{-i,j})|\alpha>&=\hat{\pi}(F_{-i,j})|\alpha>=0\quad
                             \hbox{\rm for}\ i<j,\cr
         \pi(F_{-i,i})|\alpha>&=-{(-)^i\over 2}|\alpha>\quad \hbox{\rm for}\
i>0,\cr
         \hat{\pi}(F_{-i,i})|\alpha>&=0.\cr}
\forno$$
Clearly $V_\alpha$ is also a highest weight module for $BW_{1+\infty}$, viz
$$\hat\pi\cdot \rho (W_k(f))=\sum_{j\in {\bf
Z}}(-)^{k+j}f(-j):\phi_{k-j}\phi_j:,
\forno$$
and
$$\hat\pi\cdot \rho(W_k(f))|\alpha>=0\quad\hbox{\rm for }k\ge
0.\forno$$
{}From now on we will omit $\oi$, $\hat\pi$ and $\hat\pi\cdot\rho$,
whenever no confusion can arise.

\beginsection \chapno . Vertex operators\par
\noindent\secno .
Using the boson fermion correspondence (see e.g. [DJKM 3], [K], [tKL] and [Y]
),
we can express the
fermions in terms of differential operators, i.e. there exists an
isomorfism $\sigma:V\to {\bf C}[\theta,t_1,t_3,\cdots]$, where $\theta^2=0$,
$t_it_j=t_jt_i$, $\theta t_j=t_j\theta$ and
$V_{\alpha}=\theta^{\alpha}{\bf C}[t_1,t_3,\cdots].$
such that $\sigma(|0>)=1$.
Define the following two generating series (fermionic fields):
$$\phi^{\pm}(z)=\sum_{j\in{\bf Z}} \phi_j^{\pm} z^{-{j}}
               =\sum_{j\in{\bf Z}}(\pm)^j\phi_j z^{-j}, \forno
$$
then one has the following vertex operator for these fields:
$$\sigma\phi^{\pm}(z)\sigma^{-1}=
{\theta+{\partial\over\partial\theta}\over\sqrt 2}
\exp(\pm\sum_{j>0,{\rm odd}}t_jz^{j})
\exp(\mp2\sum_{j>0,{\rm odd}}{\partial\over\partial t_j}{z^{-{j}}\over j}).
\forno$$
\secno .
Define
$$\eqalign{W(y,w)&=\sum_{\ell=0}^\infty {(y-w)^\ell\over \ell
!}W^{(\ell+1)}(w)\cr
&=\sum_{\ell=0}^\infty {(y-w)^\ell\over \ell !}\sum_{k\in{\bf
Z}}W_k^{(\ell+1)}w^{-k-\ell-1}\cr
&:={:\phi^+(y)\phi^-(w):\over w},}\forno$$
then
$$W^{(\ell+1)}(z)=:{\partial^\ell\phi^+(z)\over\partial
z^\ell}{\phi^-(z)\over z}:\forno$$
and
$$\eqalign{W^{(\ell+1)}_k&=W_k(-\ell! {D\choose \ell})\cr
&=-t^kD(D-1)\cdots
(D-\ell+1)+(-D)(-D-1)\cdots(-D-\ell+1)(-t)^k\cr
&=-t^{k+\ell}({\partial\over\partial t})^\ell
+(-)^{k+\ell}t({\partial\over\partial t})^\ell t^{k+\ell-1}.}\forno$$
Using (4.2), we find that for $|w|<|y|$
$$W(y,w)={1\over 2}{y+w\over y-w}(X(y,w)-w^{-1}),\forno$$
where $X(y,w)$ is the vertex operator defined in (1.3). Hence,
$$W^{(\ell)}(z)={w\over\ell}{\partial^\ell X(y,z)\over\partial z^\ell}|_{y=z}
+{1\over 2}{\partial^{\ell-1} X(y,z)-z^{-1}\over\partial
z^{\ell-1}}|_{y=z}.
\forno$$
Define
$$\alpha_j(z)=\cases{{1\over 2}x&if $j=-1$,\cr
-{j\over 2}t_{-j}& if $j<2$ odd,\cr
{\partial\over\partial x}&if $j=1$,\cr
{\partial\over\partial t_j}&if $j>2$ odd,\cr}\forno$$
and their generating series by
$$\alpha(z)=\sum_{j\in {\bf Z}}\alpha_jz^{-j-1},\forno$$
then $[\alpha_j, \alpha_k]={j\over 2}\delta_{j,-k}$. Since
$X(z,z)=z^{-1}$,
one finds the following expression for $W^{(\ell)}(z)$:
$$W^{(\ell)}(z)={2\over\ell}:(2\alpha(z)+{\partial\over\partial
z})^{\ell-1}\alpha(z):+{1\over z}:(2\alpha(z)+{\partial\over\partial
z})^{\ell-2}\alpha(z):.\forno$$
For $\ell=1,2,3$ one finds respectively
$$\eqalign{W^{(1)}(z)&=2\alpha(z),\cr
W^{(2)}(z)&=2:\alpha(z)^2:+{\partial\alpha(z)\over\partial
z}+{\alpha(z)\over z},\cr
W^{(3)}(z)&={8\over 3}:\alpha(z)^3:
+{8\over 3}:\alpha(z){\partial\alpha(z)\over\partial z}:+{2\over
z}:\alpha(z)^2:
+{\partial\alpha(z)\over\partial
z}+{2\over 3}{\partial^2\alpha(z)\over\partial
z^2}.\cr}$$

\beginsection \chapno . The BKP hierarchy \par
\secno .
The BKP hierarchy is the following equation for
$\tau=\tau(t_1,t_3,\ldots)$ (see e.g [DJKM3], [K], [L2], [Y]):
$${\rm Res}_{z=0} {dz\over z} \phi^+(z)\tau\otimes\phi^-(z)\tau=
{1\over 2}\theta \tau\otimes \theta\tau.
\forno$$
Here ${\rm Res}_{z=0} dz \sum_j f_jz^j=f_{-1}$.
We assume that $\tau$ is any solution of (5.1), so we no longer
assume that $\tau$ is a polynomial in $t_1,t_3,\ldots$.

We proceed now to rewrite (5.1) in terms of formal
pseudo--differential operators. We start by multiplying (5.1) from the
left with ${\partial\over\partial \theta}\otimes{\partial\over\partial
\theta}$ and divide both the first and the last component of the
tensor product by $\tau(t)$. Let $x=t_1$ and
$\partial={\partial\over\partial x}$, then (5.1) is equivalent to
the following bilinear identity:
$${\rm Res}_{z=0} {dz\over z}w(x,t,z)w(x',t',-z)=1,
\forno$$
where
$$\eqalign{
w(x,t,\pm z)&=W(x,t,\pm
z)e^{xz}=W(x,t,\partial)e^{xz}\qquad\hbox{with}\cr
W(x,t,z)&=P(x,t,z)e^{\xi(t,z)},\quad \xi(t,z)=\sum_{i\ge
3}t_iz^i\qquad\hbox{and}}\forno$$
$$\eqalign{P(x,t, z)
&={e^{-\eta(x,t,z)}\tau(x,t)
   \over\tau(x,t)}\cr
&={\tau(x_-{2\over z},t_3-{2\over 3z^3},t_5-{2\over 5z^5},\cdots)\over
\tau(x,t)}
 =:{\tilde\tau(x,t,z)\over\tau(x,t)},\cr}
\forno$$
where $\eta(x,t,z)=2({\partial\over\partial
x}z^{-1}+\sum_{j>2}{\partial\over\partial t_j}{z^{-j}\over j})$,
for convenience we also define $\xi (x,t,z)=\xi(t,z)e^{xz}$.

\noindent \secno . As usual one denotes the differential part of a
pseudo--differential
operator $P=\sum_j P_j\partial^{-j}$ by $P_+=\sum_{j\ge 0}
P_j\partial^{-j}$ and writes $P_-=P-P_+$. The anti--involution $*$ is
defined as follows $(\sum_j P_j\partial^{-j})^*=
\sum_j (-\partial)^{-j}P_j$
One has the following fundamental lemma.
\proclaim Lemma \prono .
Let $P(x,t,\partial)$ and $Q(x,t,\partial)$ be two formal
pseudo--differential operators, then
$$(P(x,t,\partial)Q(x,t',\partial))_-=\pm
\sum_{i>0}R_i(x,t,t')\partial^{-i}$$
if and only if
$${\rm Res}_{z=0}{dz}P(x,t,\partial)e^{\pm xz}
Q(x',t',\partial)e^{\mp x'z}=
\sum_{i>0}R_i(x,t,t'){(x-x)^{i-1}\over(i-1)!}.$$

\noindent The proof of this lemma is analogous to the proof of Lemma
4.1 of [L1] (see also [KL]).

\noindent \secno . Now differentiate (5.2) to $t_k$, where we assume
that $x=t_1$,
then we obtain
$${\rm Res}_{z=0}{dz\over z}({\partial P(x,t,z)\over\partial t_k}+P(x,t,z)z^k)
e^{\xi(x,t,z)}P(x't',-z)e^{-\xi(x',t',z)}=0.
\forno$$
Now using lemma 5.1 we deduce that
$$(({\partial P\over\partial t_k}+P\partial^k)\partial^{-1}P^*)_-=0.$$
{}From the case $k=1$ we then deduce that $P^*=\partial P^{-1}\partial^{-1}$,
if $k\ne 1$, one thus obtains
$${\partial P\over\partial x_k}= -(P\partial^kP^{-1}\partial^{-1})_-\partial P.
\forno$$
Since $k$ is odd, $\partial^-1(P\partial^kP^{-1})^*\partial=-P\partial^kP^{-1}$
and hence
$(P\partial^kP^{-1}\partial^{-1})_-\partial=(P\partial^kP^{-1})_-$. So
(5.6)
turns into Sato's equation:
$${\partial P\over \partial t_k}=-(P\partial^kP^{-1})_-P.\forno$$
\secno .
Define the operators
$$\eqalign{L&=W\partial W^{-1}=P\partial P^{-1},\quad \Gamma=x+\sum_{j>2}
jt_j\partial^{j-1},\cr
M=&WxW^{-1}=P\Gamma P^{-1}\quad\hbox{\rm and }N=ML.\cr}\forno$$
Then $[L,M]=1$ and $[L,N]=L$. Let $B_k=(L^k)_+$, using (5.7) one deduces
the following Lax equations:
$${\partial L\over \partial t_k}=[B_k,L],\quad
         {\partial M\over\partial t_k}=[B_k,M]\quad\hbox{and }
{\partial N\over\partial t_k}=[B_k,N].
\forno$$
The first equation of (5.9) is equivalent to the following Zakharov Shabat
equation:
$${\partial B_j\over\partial t_k}-{\partial B_k\over\partial t_j}=[B_k,B_j],
\forno$$
which are the compatibility conditions of the following linear problem for
$w=w(x,t,z)$:
$$Lw=zw, \quad Mw={\partial w\over\partial z}\quad\hbox{\rm and}\quad
{\partial w\over\partial t_k}=B_kw.\forno$$
\secno .
The formal adjoint of the wave function $w$ is (see [DJKM]):
$$w^*=w^*(x,z)=P^{*-1}e^{-\xi(x,t,z)}
                     =\partial P\partial^{-1}e^{-\xi(x,t,z)}.\forno$$
Now $L^*=-\partial L\partial^{-1}=-\partial P\partial P^{-1}\partial^{-1}$
and $M^*=\partial P\partial^{-1}\Gamma\partial P^{-1}\partial^{-1}$, so
$[L^*,M^*]=-1$ and
$$L^*w^*=zw^*, \quad M^*w^*=-{\partial w^*\over\partial z}\quad\hbox{\rm
and}\quad
{\partial w^*\over\partial x_k}=-(L^{*k})_+w^*=-B^*_kw^*.\forno$$
Finally, notice that by differentiating the bilinear identity (5.2) to $x_1'$
we obtain
$${\rm Res}_{z=0}dz w(x,t,z)w^*(x',t',z)=0.\forno$$

\beginsection \chapno . Proof of Theorem 1\par
\secno . In this section we prove Theorem 1.1. We start from the
bilinear identity (5.14)  and multiply it by
$\tau(x,t)$, which gives
$${\rm Res}_{z=0}dz e^{-\eta(x,t,z)}\tau(x,t)e^{\xi(x,t,z)}
{\partial\over\partial x'}\left ({e^{\eta(x',t',z)}\tau(x',t')
\over \tau(x',t')}e^{-\xi(x',t',z)}\right )=0.\forno$$
Now let $(1-w/y)^{-1}(1+w/y)X(y,w)$ act on this identity, then one
obtains
$$\eqalign{&{\rm Res}_{z=0}{dz\over wz} {1+w/y\over 1-w/y}{1-z/y\over
1+z/y}{1+z/w\over 1-z/w}
e^{-\eta(x,t,z)-\eta(x,t,y)+\eta(x,t,w)}\tau(x,t)e^{\xi(x,t,z)+\xi(x,t,y)-\xi(x,t,w)}
\times\cr
&\qquad {\partial\over\partial x'}\left ({e^{\eta(x',t',z)}\tau(x',t')
\over \tau(x',t')}e^{-\xi(x',t',z)}\right )=0.\cr}\forno$$
Next use the fact that
$(1-u)^{-1}(1+u)=2\delta(u,1)-(1-u^{-1})^{-1}(1+u^{-1})$,
where $\delta(u,v)=\sum_{j\in{\bf Z}} u^{-j}v^{j-1}$, then (6.2) is
equivalent to
$$\eqalign{&-{\rm Res}_{z=0}{dz\over wz} {1+w/y\over 1-w/y}{1-y/z\over
1+y/z}{1+w/z\over 1-w/z}
e^{-\eta(x,t,z)-\eta(x,t,y)+\eta(x,t,w)}\tau(x,t)e^{\xi(x,t,z)+\xi(x,t,y)-\xi(x,t,w)}
\times\cr
&\qquad {\partial\over\partial x'}\left ({e^{\eta(x',t',z)}\tau(x',t')
\over \tau(x',t')}e^{-\xi(x',t',z)}\right )=
{2\over w}(e^{-\eta(x,t,y)}\tau(x,t)e^{\xi(x,t,y)}
{\partial\over\partial x'}\left ({e^{\eta(x',t',w)}\tau(x',t')e^{-\xi(x',t',w)}
\over \tau(x',t')}\right )\cr
&\qquad -e^{\eta(x,t,w)}\tau(x,t)e^{-\xi(x,t,w)}
{\partial\over\partial x'}\left ({e^{-\eta(x',t',y)}\tau(x',t')
\over \tau(x',t')}e^{\xi(x',t',y)}\right )).\cr}\forno$$
Divide this formula by $\tau(x,t)$, then it turns into
$$\eqalign{&-{\rm Res}_{z=0}{dz\over z}e^{-\eta(x,t,z)}\left (
{1+w/y\over1-w/y}
{X(y,w)\tau(x,t)\over\tau(x,t)}\right )
{e^{-\eta(x,t,z)}\tau(x,t)\over \tau(x,t)}e^{\xi(x,t,z)}
{\partial\over\partial x'}\left ({e^{\eta(x',t',z)}\tau(x',t')e^{-\xi(x',t',z)}
\over \tau(x',t')}\right )\cr
&=2{\rm Res}_{z=0}{dz\over z}\delta(w,z) (
{e^{-\eta(x,t,y)}\tau(x,t)\over \tau(x,t)}e^{\xi(x,t,y)}
{\partial\over\partial x'}\left ( {e^{\eta(x',t',z)}\tau(x',t')
\over \tau(x',t')}e^{-\xi(x',t',z)}\right )\cr
&\qquad -
{e^{\eta(x,t,z)}\tau(x,t)\over \tau(x,t)}e^{-\xi(x,t,z)}
{\partial\over\partial x'}\left ( {e^{-\eta(x',t',y)}\tau(x',t')
\over \tau(x',t')}e^{\xi(x',t',y)}\right ) )\cr
&=2{\rm Res}_{z=0}{dz\over w^2}\sum_{\ell=0}^\infty {(y-w)^\ell\over
\ell !}\sum_{k\in{\bf Z}}( ({z\over w}
)^{k+\ell-1}({\partial\over\partial z})^\ell (W(x,t,z)e^{xz})
{\partial\over\partial x'}(W(x',t',-z)e^{-x'z})\cr
&\qquad -
W(x,t,-z)e^{-xz}
({z\over w}
)^{k+\ell-1}({\partial\over\partial z})^\ell{\partial\over\partial x'}
(W(x',t',z)e^{x'z}) )\cr
&=2{\rm Res}_{z=0}{dz}\sum_{\ell=0}^\infty {(y-w)^\ell\over
\ell !}\sum_{k\in{\bf Z}}w^{-k-\ell-1} (
W(x,t,z)x^\ell\partial^{k+\ell-1}e^{xz}{\partial\over\partial x'}
(W(x',t',-z)e^{-x'z})\cr
&\qquad -W(x,t,-z)e^{-xz}{\partial\over\partial x'}
(W(x',t',z)x^\ell\partial^{k+\ell-1}e^{x'z})).\cr}
\forno$$
Now define
$$\sum_{j=0}^\infty c_j(x,t,y,w)z^{-j}=e^{-\eta(x,t,z)}\left (
{1+w/y\over1-w/y}
{X(y,w)\tau(x,t)\over\tau(x,t)}\right ),\forno$$
then the first line of (6.4) is equal to
$$-{\rm Res}_{z=0}{dz}\sum_{j=0}^\infty c_j(x,t,y,w)L^{-j-1}
(W(x,t,z)e^{xz})
{\partial\over\partial x'}(W(x',t',-z)e^{-x'z}).$$
Now using Lemma  5.1 with $t=t'$, one deduces that
$$\eqalign{{1\over 2}&
\sum_{j=1}^\infty c_j(x,t,y,w)L^{-j}=\cr
&-
\sum_{\ell=0}^\infty {(y-w)^\ell\over
\ell !}\sum_{k\in{\bf Z}}w^{-k-\ell-1} (
W(x,t,z)x^\ell\partial^{k+\ell}W(x,t,z)^{-1}-(-)^{k+\ell}W(x,t,z)
\partial^{k+\ell-1}x^\ell\partial W(x,t,z)^{-1})_- .}\forno$$
So finally one has
$$\eqalign{{1\over 2}&(e^{-\eta(x,t,z)}-1)\left ( {1+w/y\over1-w/y}
{X(y,w)\tau(x,t)\over\tau(x,t)}\right )w(x,t,z)=\cr
& -\sum_{\ell=0}^\infty {(y-w)^\ell\over
\ell !}\sum_{k\in{\bf Z}}w^{-k-\ell-1} (M^\ell L^{k+\ell}
-(-)^{k+\ell}L^{k+\ell-1}M^\ell L)_-w(x,t,z),\cr}$$
wich is equal to the Adler--Shiota--van Moerbeke formula (1.2) for the
BKP case:
$$(w+y)(e^{-\eta(x,t,z)}-1)\left (
{X(y,w)\tau(x,t)\over\tau(x,t)}\right )w(x,t,z)=
2(w-y)Y(y,w)_- w(x,t,z).$$
\noindent \secno . Since the left--hand--side of (1.2) is also equal to
$$(w+y)(e^{-\eta(x,t,z)}-1)\left (
{(X(y,w)-1)\tau(x,t)\over\tau(x,t)}\right )w(x,t,z),$$
we have the following corollary of Theorem 1.1:
\proclaim Corollary \prono . For $k\in{\bf Z}$ and $f$ some polynomial
one has
$${\sigma \cdot\hat\pi\cdot\rho(W_k(f))\tau(x,t)\over \tau(x,t)}=
{(f(N)L^k-(-L)^kf(-N))_-w(x,t,z)\over w(x,t,z)}.\forno$$

\vskip .3cm
\noindent{\bf References} \vskip .3cm
\halign{#\hfil&\quad\vtop{\parindent=0pt\hsize=41em\strut#\strut}\hfill\cr
[{\bf ASV1} ]&{ M. Adler, T. Shiota and P. van Moerbeke,
{}From the $w_\infty$-algebra to its central extension: a
$\tau$-function approach,  to appear in Physics Letters A.}\cr

[{\bf ASV2} ]&{M. Adler, T. Shiota and P. van Moerbeke,
A Lax representation of the vertex operator and the central extension,
to appear in Comm. Math. Phys.}\cr

[{\bf DJKM1} ]&{E. Date, M. Jimbo, M. Kashiwara and T. Miwa,
Transformation groups for soliton equations.  Euclidean Lie
algebras and reduction of the KP hierarchy , Publ. Res. Inst.
Math. Sci. {\bf  18} (1982),  1077--1110.}\cr

[{\bf DJKM2} ]&{E. Date, M. Jimbo, M. Kashiwara and T. Miwa,
Transformation groups for soliton equations , in:
Nonlinear integrable systems---classical theory and quantum theory
eds M. Jimbo and T. Miwa, World Scientific, 1983),  39--120.}\cr

[{\bf DJKM3} ]&{E. Date, M. Jimbo, M. Kashiwara and T. Miwa,
Transformation groups for soliton equations IV. A new hierarchy of
soliton equations of KP type, Physica 4D (1982), 343--365.}\cr

[{\bf K} ]&{V.G. Kac , Infinite dimensional Lie algebras,
Progress in Math., vol. 44, Brikh\"{a}user, Boston, 1983; 2nd
ed., Cambridge Univ. Press, 1985; 3d ed., Cambridge Univ. Press,
1990.}\cr

[{\bf KL}]&{V. Kac and J. van de Leur, The $n$--Component KP hierarchy and
Representation Theory., in Important Developments in Soliton Theory, eds. A.S.
Fokas and V.E. Zakharov. Springer Series in Nonlinear Dynamics, (1993),
302--343.}\cr

[{\bf KP} ]&{V.G. Kac and D.H. Peterson , Spin and wedge
representations of infinite dimensional Lie algebras and groups
, Proc. Nat. Acad. Sci U.S.A. (1981),  3308--3312.
}\cr

[{\bf KR}]&{V. Kac and A. Radul, Quasifinite highest weight
modules over the Lie algebra of differential operators on the circle
, Comm. Math. Phys. {\bf  157 } (1993),  429-457.
}\cr

[{\bf tKL} ]&{F. ten Kroode and J. van de Leur,
Level one representations of the
twisted affine algebras $A_n^{(2)}$ and $D_n^{(2)}$, Acta Appl. Math.
{\bf 27} (1992), 153 -- 224.
}\cr

[{\bf L1} ]&{J. van de Leur, The $W_{1+\infty}(gl_s)$--symmetries of
the $s$--component KP hierarchy. hep-th
9411069.}\cr

[{\bf L2} ]&{J. van de Leur, The $n$--th reduced BKP hierarchy, the
string equation and $BW_{1+\infty}$--constraints. hep-th
9411067.}\cr

[{\bf OS} ]&{A.Yu. Orlov and E.I. Schulman, Additional
symmetries for integrable and conformal algebra representations,
Lett. Math. Phys. 12 (1986),  171-.}\cr

[{\bf R} ]&{A.O. Radul, Lie algebras of differential operators,
their central extensions, and W--algebras , Funct. Anal. and its Appl.
{\bf  25
} (1991),  33--49.}\cr

[{\bf Y} ]&{Y.--C. You, Polynomial solutions of the BKP hierarchy
and projective representations of  symmetric groups, in:
Infinite dimensional Lie algebras and groups, ed. V.G. Kac,
Adv, Ser. in
Math. phys. 7, world Sci., 1989, 449--466.}\cr}

\vskip .5cm
\noindent
{\bf JOHAN VAN DE LEUR}

\noindent FACULTY OF APPLIED MATHEMATICS

\noindent UNIVERSITY OF TWENTE

\noindent P.O. BOX 217

\noindent 7500 AE ENSCHEDE

\noindent THE NETHERLANDS

\end